\documentclass[journal]{IEEEtran}
\IEEEoverridecommandlockouts
\usepackage{cite}
\usepackage{amsmath,amssymb,amsfonts}
\usepackage{amsthm} 
\usepackage{algorithmic}
\usepackage{graphicx}
\usepackage{textcomp}
\usepackage{amsfonts} 
\usepackage{xcolor}     
\usepackage{balance}    
\usepackage{booktabs}   
\usepackage{amsthm}     
\usepackage{floatrow}
\usepackage{subfig}
\usepackage[capitalize]{cleveref}
\usepackage{caption} %
\usepackage[font=small]{caption} 
\usepackage{enumerate}
\newtheorem{remark}{Remark} 
\newtheorem{theorem}{Theorem}

\newtheorem{proposition}{Proposition}

\begin{document}
	
	\title{Pinching-Antenna Systems For Indoor Immersive Communications: A 3D-Modeling Based Performance Analysis}
	
	\author{Yulei Wang, 
		Yalin Liu, 
		Yaru Fu, 
		and Zhiguo Ding,~\IEEEmembership{Fellow,~IEEE}
		
		\thanks{Y. Wang is with the Hubei Key Laboratory of Intelligent Wireless Communications, Hubei Engineering Research Center of Intelligent Internet of Things Technology, College of Electronics and Information Engineering, South-Central Minzu University, Wuhan 430074, China (e-mail: ylwang@mail.scuec.edu.cn).}
		
		\thanks{Y. Liu and Y. Fu are with the School of Science and Technology, Hong Kong Metropolitan University, Hong Kong (e-mail: ylliu@hkmu.edu.hk; yfu@hkmu.edu.hk).}
		
		\thanks{Z. Ding is with the University of Manchester, Manchester, M1 9BB, UK, and Khalifa University, Abu Dhabi, UAE (e-mail: zhiguo.ding@ieee.org).}
	}

	\markboth{Journal of \LaTeX\ Class Files,~Vol.~14, No.~8, August~2025}%
	{Shell \MakeLowercase{\textit{et al.}}: Bare Demo of IEEEtran.cls for IEEE Journals}
	
	
	\maketitle
	
	\begin{abstract}
		The emerging pinching antenna (PA) technology has high flexibility to reconfigure wireless channels and combat line-of-sight blockage, thus holding transformative potential for indoor immersive applications in 6G. This paper investigates Pinching-antenna systems (PASS) for indoor immersive communications. Our contributions are threefold: (1) we construct a 3D model to characterize the distribution of users, waveguides, and PAs in the PASS; (2) we develop a general theoretical model on downlink performance of PASS by capturing 
		PA-user relationships and system parameters' impacts; and (3) we conduct comprehensive numerical results of the theoretical model and provide implementation guidelines for PASS deployments.
	\end{abstract}
	
	
	\IEEEpeerreviewmaketitle
	
	\section{Introduction}
\begin{figure}[t]
		\centering
		\includegraphics[scale=0.5]{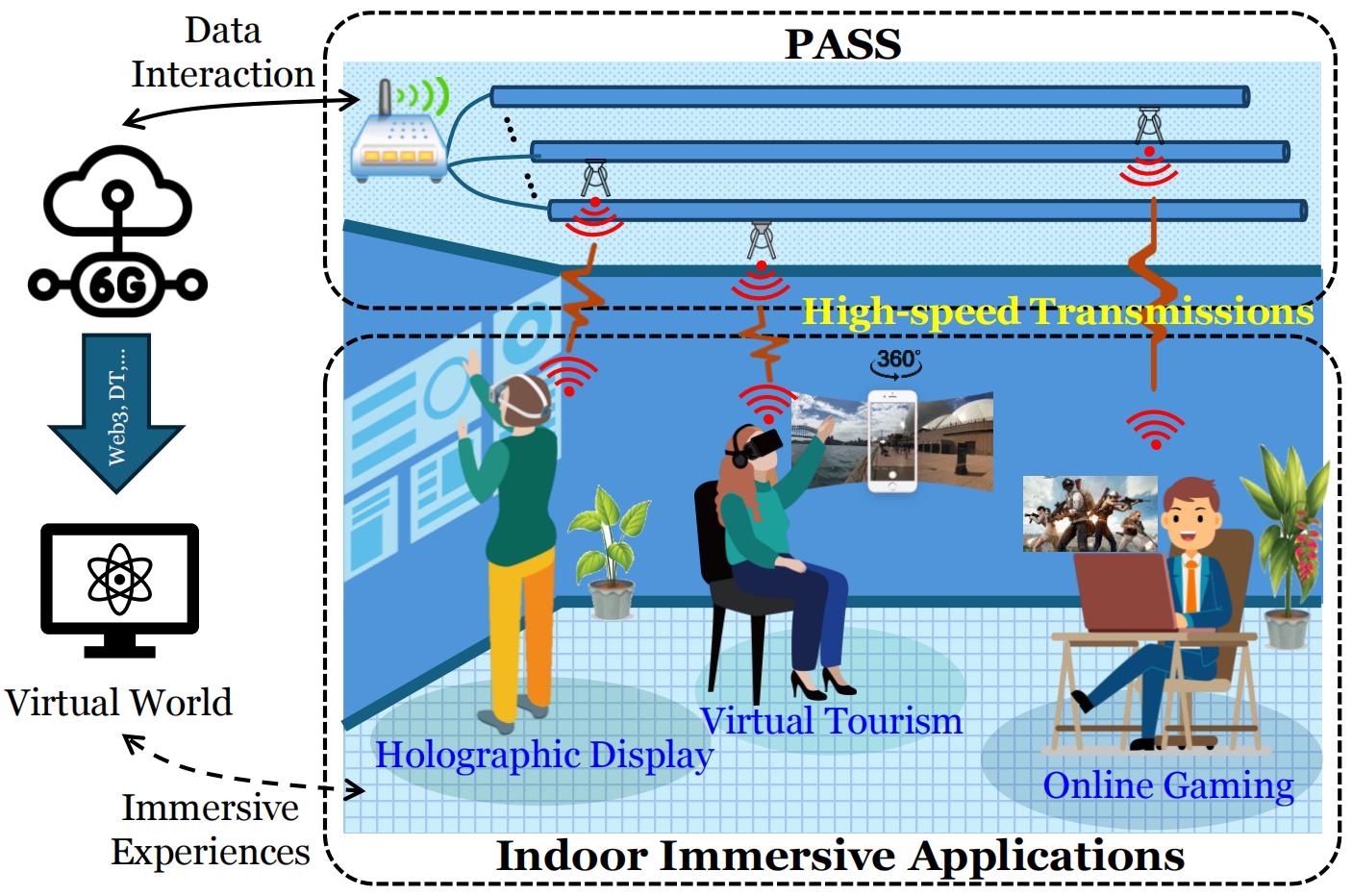}
		\caption{PASS for indoor immersive communications.}
		\label{Fig-Pinching-Overview}
	\end{figure}

	\IEEEPARstart{T}{he} upcoming 6G networks unlock a new era of immersive communications by seamlessly integrating physical infrastructures and virtual worlds, facilitating the widespread adoption of extended reality, metaverse, and digital twin \cite{Du23AIGC, liu24digital}. As shown in~\cref{Fig-Pinching-Overview}, benefited by immersive communications, a variety of indoor applications are emerging, such as holographic displays, virtual tourism, online gaming, and intelligent manufacturing. These indoor immersive applications will fundamentally revolutionize the way people work, entertain, and communicate through lifelike interactions and immersive experiences. However, the devices used for these indoor applications need to operate at high-frequency bands to achieve high-speed data rates and real-time transmissions, making them inherently vulnerable to significant path loss and line-of-sight (LoS) from indoor obstacles. Consequently, these applications impose stringent requirements for network capacity and reliability while facing substantial communication vulnerabilities from fixed deployments of conventional wireless systems.
	
	To address the above vulnerabilities, more flexible wireless technologies are required to improve network performance for indoor immersive applications. Pinching antennas (PAs), first demonstrated by NTT DOCOMO in 2021 \cite{suzuki22pinching}, represent a groundbreaking technology that is capable of reconfiguring wireless channels and establishing LoS connections flexibly~\cite{ding25flexible, liu25Tutorial}. This remarkable flexibility positions the pinching-antenna system (PASS) as a promising solution for indoor immersive applications. 
	An indoor PASS deployment is illustrated in~\cref{Fig-Pinching-Overview}. With high flexibility, PAs are along line-shaped waveguides, adaptively controlled and positioned to serve users for overcoming the transmission vulnerabilities from indoor obstacles. Although prior work has explored various applications of PA, such as PA-assisted integrated sensing and communications (ISAC) \cite{qin25ISAC,zhang25ISAC}, non-orthogonal multiple access integration (NOMA) \cite{fu25NOMA}, and PA-assisted simultaneous wireless information and power transfer (SWIPT) \cite{Wang25NOMA}, the critical domain of PA-assisted indoor immersive communications remains unexplored. This gap, coupled with the absence of an accurate and generalizable performance analysis model for indoor immersive applications, motivates our work.
	
	This paper investigates PASS-enabled downlink transmission performance for indoor immersive communications. Our key contributions are threefold: \textit{\textbf{First}}, we construct a 3D system model for PASS in indoor scenarios, where users (i.e., diverse devices and sensors for immersive applications) are uniformly distributed in the serving area with each waveguide on the ceiling and served by the nearest PA along the waveguide. \textit{\textbf{Second}}, we develop a general theoretical model for analyzing the downlink performance, i.e., successful transmission probability (STP), which (i) captures the PA-user location correlations, (ii) incorporates critical system parameters (deployment settings and transmission configurations), and (iii) analyzes impacts of the PASS parameters. \textit{\textbf{Third}}, we conduct comprehensive numerical results to investigate the proposed performance model. 
	The results provide practical guidelines for optimizing system parameters to enhance the performance of PA-assisted indoor immersive communications.

	\vspace{-4 mm}
	\section{System Model}\label{Sec-Sys}
	As shown in \cref{Fig: system}(a), an access point (AP) equipped with a PASS serves users uniformly distributed in a service area, e.g., a room. Multiple dielectric waveguides of PASS are evenly installed on the ceiling to cover the whole room. Each waveguide is equipped with an adjustable PA controlled by the AP. The AP facilitates downlink transmissions for users with immersive experiences. 
	
	\vspace{-4 mm}
	\subsection{Deployment Model} 
	The service area is approximated as a rectangular cuboid with length $L$, width $D$, and height $h$. In the PASS, $2K + 1$ waveguides are installed along the wider side of the ceiling, with an inter-waveguide spacing of $d$ among each other; and each waveguide has a length equal to the room length $L$. Let $\mathbf{w} = \{w_k\}_{k \in \mathcal{K}}$ denote the set of waveguides, where $\mathcal{K} = \{0, \pm 1, \ldots, \pm K\}$. The central waveguide $(w_0)$ is positioned along the midline of the ceiling's wide side, while the remaining waveguides $(w_k,k \neq 0)$ are symmetrically arranged on either side of $w_0$. Each waveguide $w_k$ is connected to the AP via a dedicated radio frequency (RF) chain, with signal injection occurring at its feed point $f_k$.
	The users are uniformly distributed across the ground. Each user is served by its nearest waveguide. For a user $u_k$ served by the waveguide $w_k$, the position of the PA $p_k$ is the nearest location of its user, i.e., the top of the user. 
	Considering the inter-waveguide spacing $d$, the service regions of the waveguides on the ground can be modeled as multiple rectangular areas, each with a length of $L$ and a width of $d$, directly beneath the corresponding waveguides. It is worth noting that $(2K+1)d\leq D$ is required to ensure the deployment of waveguides in the room.
	
	\vspace{-4 mm}
	\subsection{3D Coordinate System}
\begin{figure}[t]
		\centering
		\includegraphics[scale=0.45]{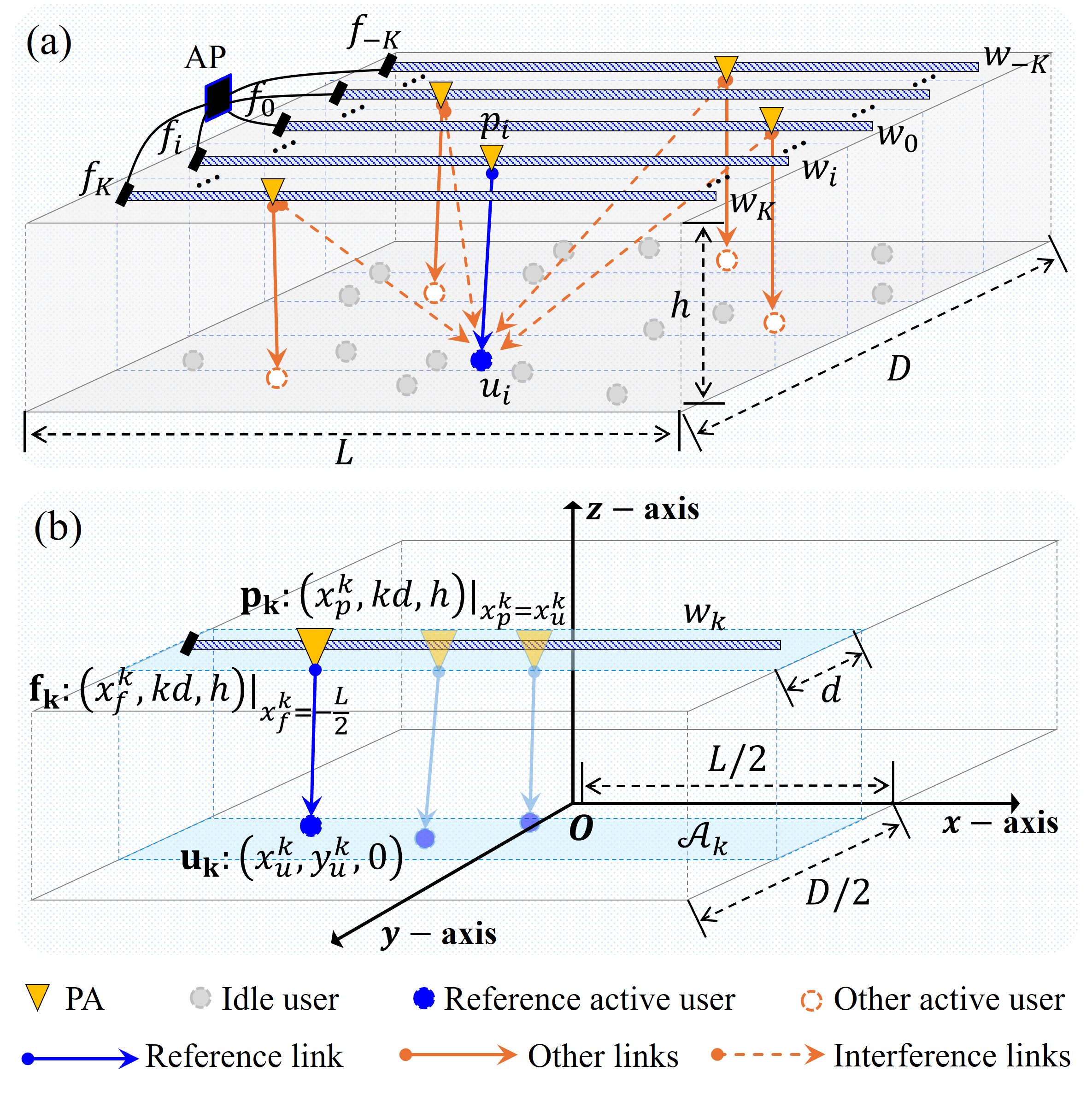}
		\caption{System model, where (a) shows the deployment of the PASS in a room and (b) shows the used $3$D coordinate system, where the blue rectangle area represents the serving area of focused $w_k$.}
		\label{Fig: system}
	\end{figure}

	We establish a 3D Cartesian coordinate system as illustrated in \cref{Fig: system}(b). The origin, denoted as $\mathbf{O}$, is located at the midpoint of the room's floor. The $xy$-plane is on the ground, the $y$-axis points toward the midpoint of the floor's wider side, the $x$-axis is 
	perpendicular to the $y$-axis and oriented in the direction opposite to the feed points of the waveguides, and the $z$-axis extends vertically toward the midpoint of the ceiling. Therefore, the position of every node in the PASS can be represented as a 3D coordinate $(x, y, z)$. For a waveguide $w_k$ positioned at $y=kd$ and $z=h$ (i.e., on the ceiling), let $\mathbf{f_k}$, $\mathbf{p_k}$, and $\mathbf{u_k}$ denote the coordinates of its feed point $f_k$, its PA $p_k$, and its served user $u_k$, respectively. We have
	\vspace{-0.3cm}
	
	{\small
		\begin{align*}
			&\mathbf{f_{k}}:\left(x_{f}^{k}, kd, h\right), \mathbf{p_{k}}:(x_p^k, kd, h),\mathbf{u_{k}}:(x_u^k, y_u^k, 0).
		\end{align*}%
	}%
	where $x_{f}^{k}=-L/2$. Since $p_k$ is movable along the waveguide $w_k$, its $x-$axis location has the value range $x_p^k\in[-L/2,L/2]$. Let $\mathcal{A}_k$ be the service region of $w_k$ on the ground, the points on $\mathcal{A}_k$ has the $x-$axis/$y-$axis values $x\in[-L/2,L/2]$ and $y\in[kd -d/2,kd+d/2]$. Given the fixed $\mathbf{u_{k}}$, $\mathbf{p_{k}}$ is adjusted to the nearest location of the served user, i.e., $x_p^k=x_u^k$. 
	
	\vspace{-4 mm}
	\subsection{Channel Model}
	To provide downlink services for users, the AP passes signals into multiple waveguides. Each downlink signal propagates through the waveguide from the feed point to the PA. The PA then radiates the signal to its served user. 
	The signal propagated through the waveguide $w_k$ experiences a phase shift~\cite{ding25flexible}. Let $h_1^k$ be the phase shift gain between the feed point $f_k$ and the PA $p_{k}$, which is given by 
	\begin{equation}
		h_1^{k}= e^{-j\frac{2\pi}{\lambda_g}\|\mathbf{f_{k}} - \mathbf{p_{k}}\|},
		\label{Eq-phase}
	\end{equation}%
	where $j$ is the imaginary unit, $\lambda_g = \lambda/n_{\text{eff}}$ is the guided wavelength with $n_{\text{eff}}$ being the effective refractive index of the dielectric waveguide \cite{ding25flexible}, $\lambda$ is the wavelength of the signal, and $\|\mathbf{f_k} - \mathbf{p_{k}}\|$ is the distance between $f_k$ and $p_{k}$. 
	
	Since the PA in each waveguide is usually activated to the top of its user, the link between them 
	is assumed as the line-of-sight (LoS) path, and the free-space channel model is used to characterize the signal propagation. Let $h_2$ be the channel between $p_{k}$ and its serving user $u_k$, which is given by
	\begin{equation}
		h_2^{k}= \frac{\sqrt{\eta}e^{-j\frac{2\pi}{\lambda}\|\mathbf{p_{k}} - \mathbf{u_k}\|}}{\|\mathbf{p_{k}} - \mathbf{u_k}\|},
		\label{Eq-loss}
	\end{equation}
	where $\eta = \lambda^2/(16\pi^2)$ is the path loss at a reference distance of 1 m, and $\|\mathbf{p_{k}} - \mathbf{u_k}\|$ is the distance between $p_{k}$ and $u_k$. Recall that $p_{k}$ and $u_k$ have the same $x-$axis location for the nearest association. Then their distance $\|\mathbf{p_{k}} - \mathbf{u_k}\|$ only depends on the height of $p_{k}$ and their $y-$axis locations.
	
	\section{Performance Modeling}\label{Sec-Ana}
	This section first provides the link model involving the transmitted signal and the received signal-to-interference-plus-noise-ratio (SINR). Accordingly, we develop the theoretical model for link performance and provide some analysis.
	
	\vspace{-4 mm}
	\subsection{Link Model}
	As shown in \cref{Fig: system}(a), we consider a typical PA-assisted downlink transmission scenario, where all waveguides simultaneously work and each serves one user. The PA in each waveguide is assumed to use the identical power, i.e., $P_t=P_{\mathrm{total}}/(2K+1)$, where $P_{\mathrm{total}}$ is the total radiation power of the AP~\cite{liu25Tutorial}. Let $s_w^{k}$ be the normalized signal fed into the waveguide $w_{k}$ (with the PA $p_{k}$), where $k \in \mathcal{K}$. 
	
	The practical resources of spectrum are limited to support the increasing number of users. Thus, interference occurs when multiple PA-assisted downlink links compete and use the same spectrum band. 
	For the reference link from the PA $p_i$ to the user $u_i$, 
	let $s_u^i$ be the received signal by $u_i$, which is given by 
	\begin{equation}
		s_u^i = \sqrt{P_t}h_1^i h_2^i s_w^i + I + w,
		\label{Eq-signal-u0}
	\end{equation}%
	where $I=\sum_{k \in \mathcal{K}/\{i\}} \alpha \sqrt{P_t}h_1^k h_2^k s_w^k$ is the interference causing by other PAs ($p_k,k \in \mathcal{K}/\{i\}$) with $\alpha$ is the interference intensity factor to accommodate these more complex scenario, e.g., various multiple access schemes and changing transmission requirements from users. 
	The term $w$ denotes the additive white Gaussian noise with zero mean and variance $\sigma^2$~\cite{ding25flexible}. 
	
	Let $\gamma_u^i$ be the SINR measured at $u_i$, which is given by $\gamma_u^i = P_u^i/(I_u^i+\sigma^2)$, 
	where $P_u^i$ is the received power of the signal from $p_i$ and $I_u^i$ is the received power of aggregative interference signals from $p_k$, which are calculated by 
	\vspace{-0.3cm}
	
	{\small
		\begin{align}
			P_u^i &= P_t |h_1^ih_2^is_w^i|^2 = \frac{ P_t } {\|\mathbf{p_i} - \mathbf{u_i}\|^2} 
			\left| e^{-2\pi j\left( \frac{\|\mathbf{f_0} - \mathbf{p_i}\|}{\lambda_g} + \frac{\|\mathbf{p_i} - \mathbf{u_i}\|}{\lambda}\right)}\right|^2 \notag  \\
			&\overset{(a)}{=} \frac{\eta P_t }{\|\mathbf{p_i} - \mathbf{u_i}\|^2},
			\label{Eq-Power}\\
			I_u^i &= \sum_{k \in \mathcal{K}/\{i\}} \eta P_t|h_1^k h_2^k s_w^k|^2=\sum_{k \in \mathcal{K}/\{i\}} \frac{\eta P_t}{\|\mathbf{p_{k}} - \mathbf{u_i}\|^2},\label{Eq-Interference}
		\end{align}%
	}%
	where \cref{Eq-Power}$(a)$ is resulted from $\left|e^{-jx} \right| = 1$~\cite{tyrovolas25Analysis}. Referring to \cref{Eq-Power,Eq-Interference}, $\gamma_u^i$ can be calculated by
		\begin{align}
			&\gamma_u^i= \frac{ \| \mathbf{p_i}-\mathbf{u_i}\|^{-2} } {\eta \sum_{k \in \mathcal{K}/\{i\}} \|\mathbf{p_k}-\mathbf{u_i}\|^{-2}  + \frac{\sigma^2}{P_t}} \notag \\
			&=\frac{ ((x_p^i-x_u^i)^2+(y_p^i-y_u^i)^2 + h^2)^{-1} }{ \eta  \sum_{k \in \mathcal{K}/\{i\}} ((x_p^k-x_u^i)^2+(y_p^k-y_u^i)^2 + h^2)^{-1} + \frac{\sigma^2}{P_t}}. \label{Eq-gamma-u0}
		\end{align}
	
	\vspace{-4 mm}
	\subsection{Tranmission Performance}
	\label{subsec: perf}
	Given a wireless link, the signal can be successfully received and decoded only when the received SINR at the receiver exceeds a given threshold. For the reference link from the PA $p_i$ to the user $u_i$, let $\gamma_0$ be the SINR threshold at $u_i$. The \textit{successful transmission probability (STP)} of the reference link, denoted by $\mathcal{P}_{s}$, is given by $\mathbb{P}\left(\gamma_u^i > \gamma_0 \right)$. Referring to \cref{Eq-gamma-u0,Eq-Power,Eq-Interference}, $\mathcal{P}_{s}$ can be evaluated by~\cref{pro1}.
	
	\begin{proposition} Given $R = \sum_{k \in \mathcal{K}/\{i\}} r_k$ and $r_k = ((x_u^k - x_u^i)^2 + (kd - y_u^i)^2 + h^2 ) ^{-1}$, the STP $\mathcal{P}_{s}$ can be evaluated by the Cumulative Distribution Function (CDF) of $R$, i.e.,
		\vspace{-0.3cm}
		
		{\small
			\begin{align*}
				&\mathcal{P}_{s}=F_{R} \left(z\right),\text{where } z= \frac{((id-y_u^i)^2 + h^2)^{-1}}{\gamma_0} - \frac{\sigma^2}{P_t}.
		\end{align*}}
		\label{pro1}
	\end{proposition}
	\vspace{-0.5cm}
	\noindent \textit{Proof.} The proof is provided in Appendix~\ref{proof_pro1}. \hfill $\blacksquare$
	
	Referring to \cref{pro1}, the calculation of $\mathcal{P}_{s}$ needs to further find $F_{R} \left(z\right)$. Recall that each PA's location is align with its served user and all users are independently distributed, and hence $F_{R} \left(z\right)$ can be derived accordingly. Thus, we have the theoretical model of $\mathcal{P}_{s}$ as given in \cref{the1}. 
	
	\begin{theorem} 
		The STP $\mathcal{P}_{s}$ of the link from $p_i$ to $u_i$ is given by
		\vspace{-0.9cm}
		
		{\small
			\begin{align*}
				& \mathcal{P}_{s} = \frac{1}{2} - \frac{1}{\pi} \left(\frac{1}{L}\right)^{2K} \int_0^{\infty} \mathrm{Im}\left(  
				\underbrace{\int_{-\frac{L}{2}}^{\frac{L}{2}} \dots \int_{-\frac{L}{2}}^{\frac{L}{2}}}_{2K} e^{jt A} \underbrace{\mathbf{d} \dots \mathbf{d}x_u^k}_{k \in \mathcal{K}/\{i\}} 
				\right) \frac{\mathbf{d}t}{t},\\
				&\text{where }A=\frac{\sigma^2}{\eta P_t}+\sum_{k \in \mathcal{K}/\{i\}}(\hat{r}_k ) ^{-1}-\frac{(\hat{r}_0)^{-1}}{\eta \gamma_0},\hat{r}_0=(id-y_u^i)^2 + h^2,\\
				&\hat{r}_k=(x_u^k - x_u^i)^2 + (kd - y_u^i)^2 + h^2, \mathrm {\mathrm{Im}}(\cdot) \text{ is the imaginary part}.
			\end{align*}%
		}%
		\label{the1}
	\end{theorem}
	\vspace{-0.5cm}
	\noindent \textit{Proof.} The proof is provided in Appendix~\ref{proof_the1}. \hfill $\blacksquare$

	\begin{remark}
		\cref{the1} leads to the following conclusions.
		
		\begin{itemize}
			\item \cref{the1} provides a general formula of STP $\mathcal{P}_{s}$ for any specific link from $p_i$ to $u_i$. This formula can explore the performance distribution given different locations of $p_i$ and $u_i$ in the room, thus giving insights for practical implementations. Particularly, \cref{Fig-Ps-u0} in \cref{Sec-Exp} shows the visualized results and specific insights.
			\item Based on \cref{the1}, $\mathcal{P}_{s}$ can be accurately calculated via two types of parameters. One is about the PASS settings  (including $2K+1$, $L$, $h$, and $d$). Another is about the transmission configurations (including $\{\gamma_0,\sigma^2,P_t,\eta \}$). 
			We give the comprehensive numerical results to analyze the impacts of these parameters in \cref{Sec-Exp}. 
			\item \cref{the1} can be easily extended to more diverse transmission scenarios, e.g., the changing values of $P_t$ and $\eta$ for different users. Accordingly, our model can be used to facilitate system optimization, e.g., dynamic power allocation of different PAs given changing transmission requirements or multiple access schemes.
		\end{itemize}
	\end{remark}

	\vspace{-4 mm}
	\section{Experimental Results}\label{Sec-Exp}
	This section presents comprehensive numerical results for the proposed theoretical model of $\mathcal{P}_{s}$ in~\cref{subsec: perf}. In our result, the carrier frequency $f_c=28$ GHz, the waveguide's refractive index $n_{\rm eff}=1.4$~\cite{ding25flexible}, and the room width $D=66$ m are used. We consider the efficient coverage strategy $(2K+1)d=D$\footnote{Thus, the interwave spacing $d$ is determined by $D/(2K+1)$.} and the equal power allocation strategy $P_t=P_{\mathrm{total}}/(2K+1)$ with $K$ and $D$ set in the particular result. Accordingly, we analyze the impacts of the following system parameters, including the reference user $u_i$'s location in the whole room, the PASS settings (i.e., $K$, $h$, $L$, and $D$), and transmission configurations (i.e., $P_{\mathrm{total}}$, $\sigma^2$, and $\gamma_0$) on $\mathcal{P}_{s}$. Considering the uniformly distributed users in the room, $10^4$ Monte Carlo simulations are implemented in MATLAB to calculate the average STP $\mathcal{P}_{s}$ and verify theoretical results. In all figures, we adopt the labels ``Theo." and ``Simu." to differentiate the theoretical and simulation results, respectively. 
	
	\begin{figure}[t]
		\captionsetup[subfigure]{justification=centering}
		\centering
		\subfloat[$\mathcal{P}_{s}$ versus $\mathbf{u_{i}}$ $(-\frac{L}{2}\leq x_u^i \leq \frac{L}{2},-\frac{D}{2}\leq y_u^i \leq \frac{D}{2})$. ]{\includegraphics[width=8.2cm]{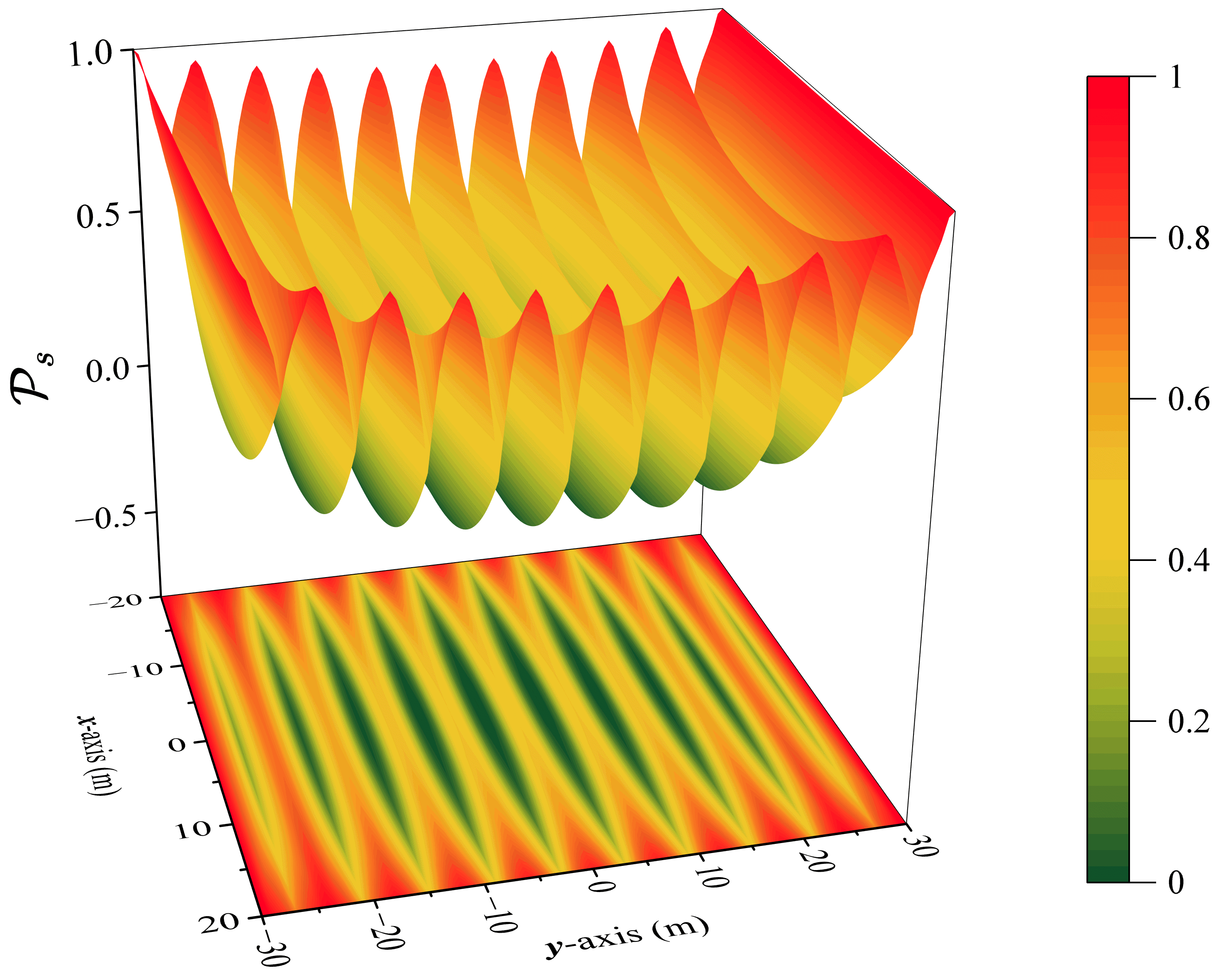}
			\label{Fig-Ps-u0-a}}	\hfil 
		\subfloat[$\mathcal{P}_{s}$ versus $\mathbf{u_{i}}$ \\
		$(-\frac{L}{2}\leq x_u^i \leq \frac{L}{2},y_u^i=0)$]{
			\includegraphics[width=4.0cm]{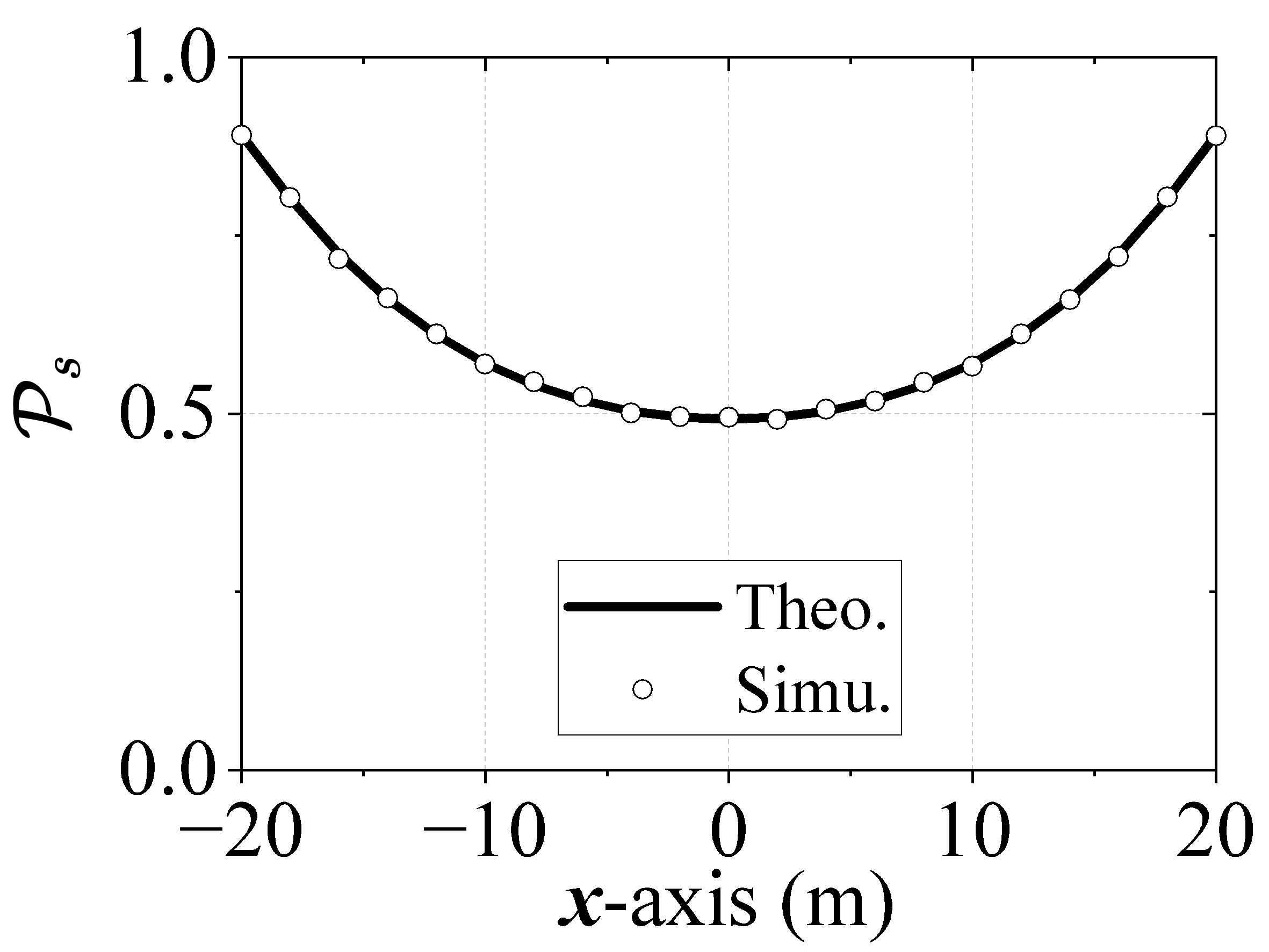}
			\label{Fig-Ps-u0-b}}	\hfil 
		\subfloat[$\mathcal{P}_{s}$ versus $\mathbf{u_{i}}$ \\
		$(x_u^i=0,-\frac{D}{2}\leq y_u^i \leq \frac{D}{2})$]{
			\includegraphics[width=4.0cm]{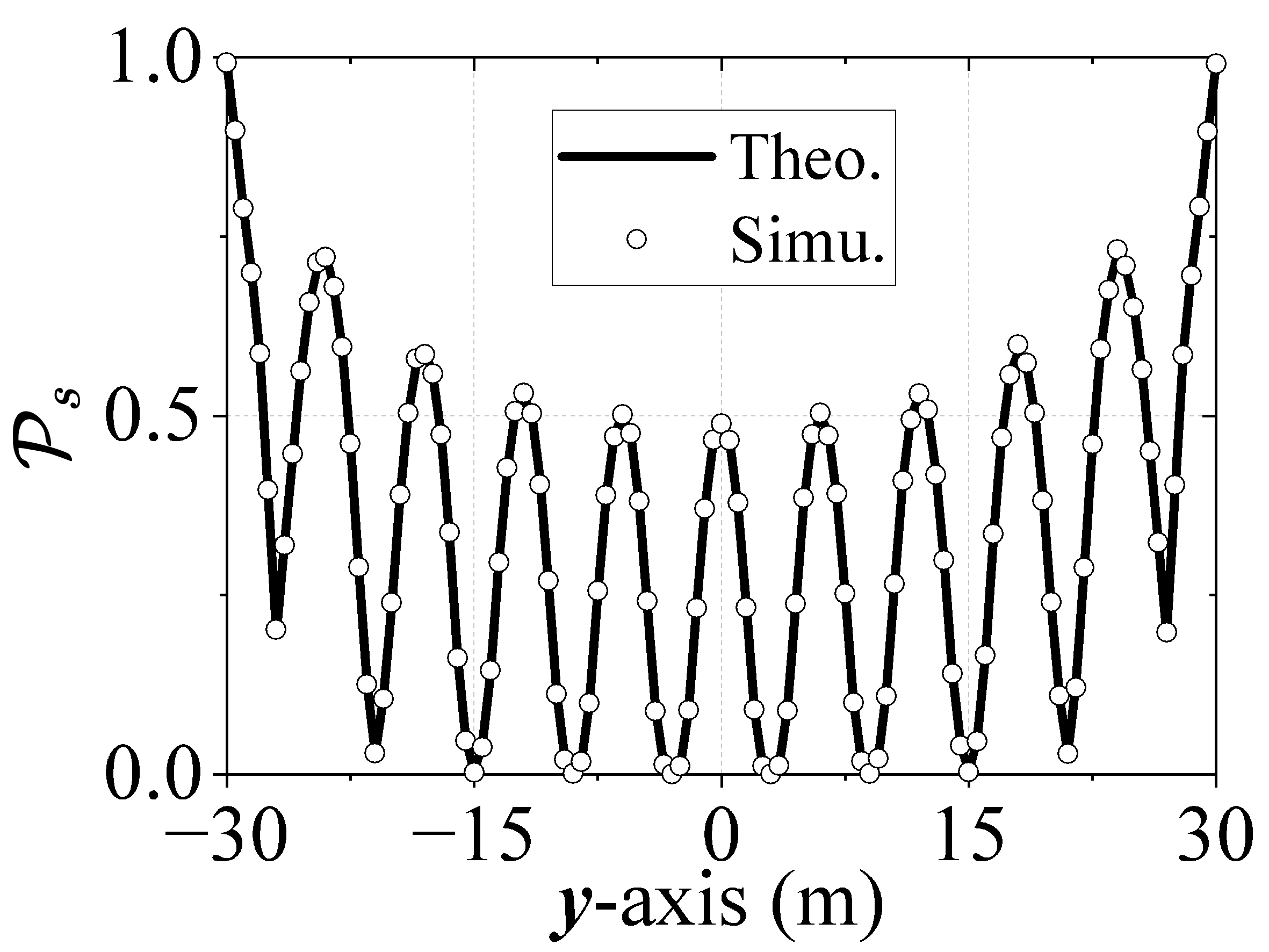}
			\label{Fig-Ps-u0-c}}	\hfil 
		\caption{$\mathcal{P}_{s}$ versus $\mathbf{u_{i}}:(x_u^i, y_u^i, 0)$ ($2K+1=11$, $h=3$ m, $L=40$ m, $P_{\mathrm{total}}=100$ dBm, $\sigma^2=-100$ dBm, $\gamma_0=2$ dB). 
		}
		\label{Fig-Ps-u0}
	\end{figure}
	
	In~\cref{Fig-Ps-u0}, we analyze the spatial distribution of $\mathcal{P}_{s}$ based on the reference user’s location, i.e., $\mathbf{u_{i}}:(x_u^i, y_u^i, 0)$, in the whole room.~\cref{Fig-Ps-u0}\subref{Fig-Ps-u0-a} plots the heatmap of the theoretical result of $\mathcal{P}_{s}$, which exhibits
	two symmetric distribution patterns. 
	\begin{itemize}
		\item \textit{Pattern 1}: Given $\mathbf{u_{i}}$ distributed in the service area $\mathcal{A}_i:([-L/2,L/2],[kd -d/2,kd+d/2], 0)$ of the waveguide $w_i$, $\mathcal{P}_{s}$'s distribution is symmetric along $w_i$ (located at $([-L/2,L/2],kd, 0)$). Particularly, $\mathcal{P}_{s}$ has the minimum value along $w_i$ and symmetrically increases by moving $\mathbf{u_{i}}$ away from $w_i$ along its two sides. 
		\item \textit{Pattern 2}: Given $\mathbf{u_{i}}$ distributed in the whole room $([-L/2,L/2],[D/2,D/2], 0)$, $\mathcal{P}_{s}$'s distribution is symmetric along the center waveguide $w_0$ (located at $([-L/2,L/2],0, 0)$). Particularly, $\mathcal{P}_{s}$ periodically varies with the changing $\mathbf{u_{i}}$ under different service areas $\{\mathcal{A}_i,i \in \mathcal{K}\}$. The overall trend is similar to \textit{Pattern 1}, where $\mathcal{P}_{s}$ has the minimum value along $w_0$ and symmetrically increases with the changing $\mathbf{u_{i}}$ far away from $w_0$. 
	\end{itemize}
	The above phenomena result from the systematic spatial distribution of interference across the service area of each waveguide. Undoubtedly, the user at the center of the service area (i.e., room) suffers the maximum interference by uniformly aggregating all interference signals. Meanwhile, the user served by different waveguides experiences the periodical distribution in the whole room because of the parallel distribution of waveguides. Additionally, \cref{Fig-Ps-u0}\subref{Fig-Ps-u0-b} and \cref{Fig-Ps-u0}\subref{Fig-Ps-u0-c} show the details of the distribution of $\mathcal{P}_{s}$ under the changing $\mathbf{u_{i}}$ along the $x$-axis and the $y$-axis, respectively. It can be seen that $\mathcal{P}_{s}$ increases when either values of $\|x_u^i\|$ or $\|y_u^i\|$ far away from the center $0$ or $id$. This is because the propagation distance of the desired signal is nearly optimized by the flexible PA's location aligning with the user, thus, the longer propagation distance of the interference signal potentially enhances the SINR. 
	
	\begin{figure}[t]
		\captionsetup[subfigure]{justification=centering}
		\centering
		\subfloat[$\mathcal{P}_{s}$ versus $K$, $h$ and $L$ \\ ($h=3$ m, $P_{\mathrm{total}}=100$ dBm, $\sigma^2=-100$ dBm, $\gamma_0=2$ dB).]{
			\includegraphics[width=8.6cm]{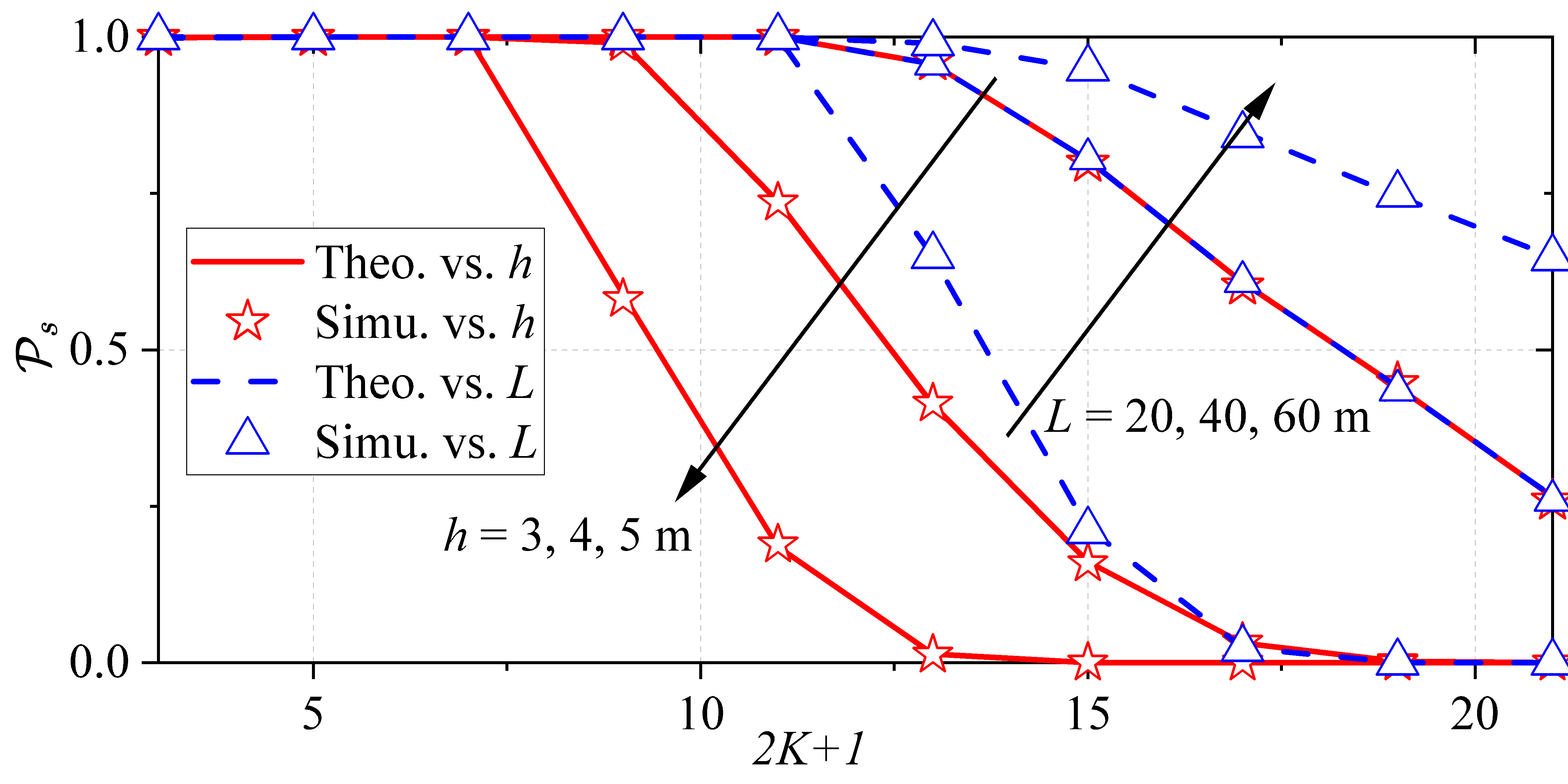}
			\label{Fig-Ps-parameter-b}}	\hfil 
		\subfloat[$\mathcal{P}_{s}$ versus $P_{\mathrm{total}}$, $\sigma^2$, and $\gamma_0$ \\ ($2K+1=11$, $h=3$ m, $L=40$ m).]{
			\includegraphics[width=8.6cm]{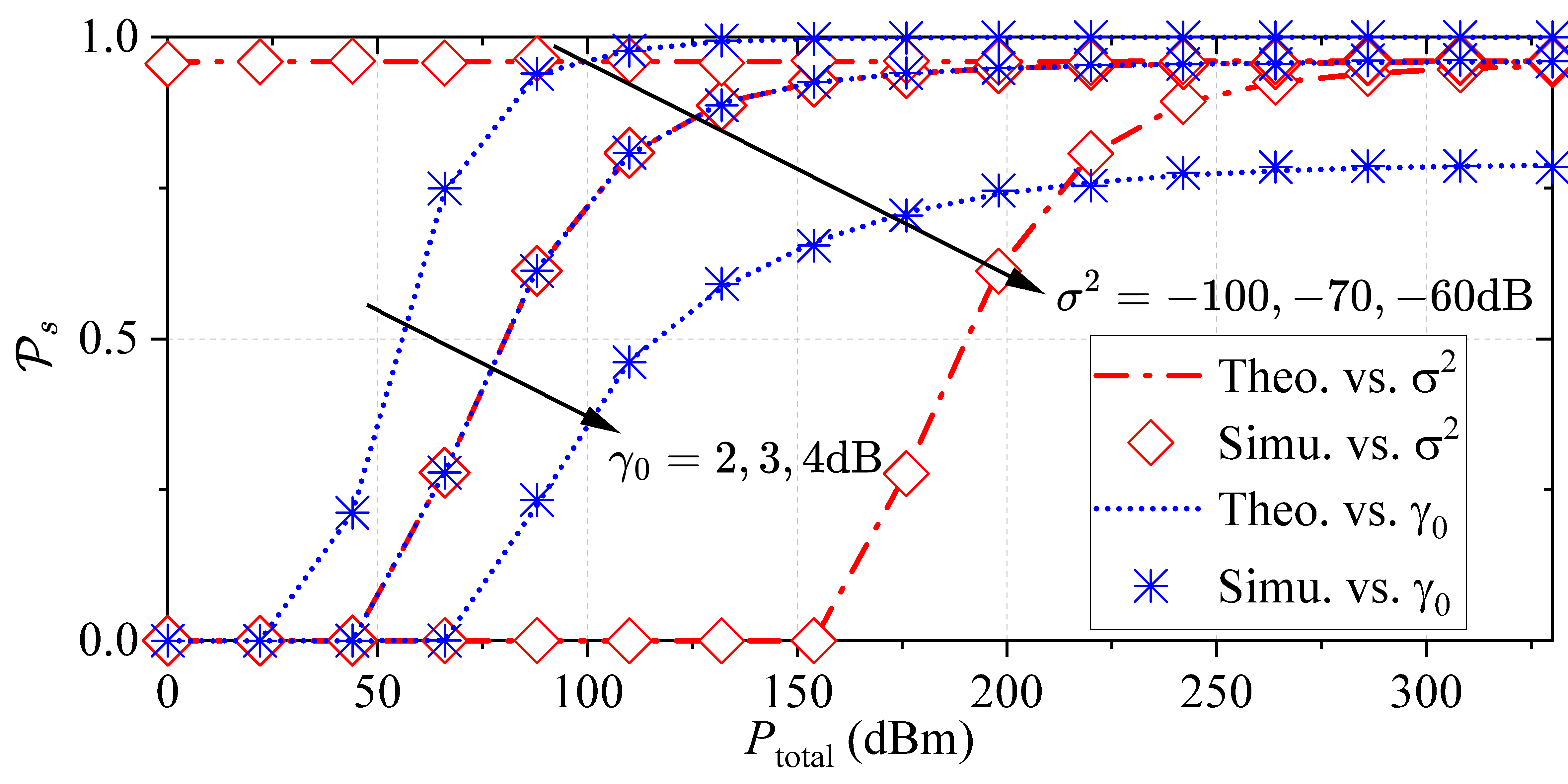}
			\label{Fig-Ps-parameter-a}}	\hfil 
		\caption{$\mathcal{P}_{s}$ versus $K$, $h$, $L$, $D$, $\gamma_0$, $P_{\mathrm{total}}$, and $\sigma^2$ ($\mathbf{u_{i}}:(0, 0, 0)$).}
		\label{Fig-Ps-parameter}
	\end{figure}
	
	In \cref{Fig-Ps-parameter}, we analyze the impacts of $K$, $h$, $L$, $D$, $\gamma_0$, $P_{\mathrm{total}}$, and $\sigma^2$ on $\mathcal{P}_{s}$. We have the following observations.
	\begin{itemize}
		\item $\mathcal{P}_{s}$ decreases as $2K+1$ increases. This is because we adopt a coverage strategy where $(2K+1)d = D$ and an equal power allocation strategy where $P_t = P_{\mathrm{total}} / (2K+1)$, with $P_{\mathrm{total}}$ and $D$ being constant. As $2K+1$ increases, both the per-waveguide power allocation and the inter-waveguide spacing decrease, while the number of interference sources grows. The latter effect reduces the SINR at the user, leading to a decrease in $\mathcal{P}_{s}$.
		\item $\mathcal{P}_{s}$ decreases as $h$ increases or $L$ increases. It occurs because a larger $h$ extends the propagation distance and lowers the received power for both desired and interference signals. Based on our setting, the power reduction of the desired signal dominates over that of the interference, leading to a final decrease in SINR and $\mathcal{P}_{s}$. Meanwhile, a larger $L$ disperses interfering PAs, reducing the aggregated interference and increasing $\mathcal{P}_{s}$.
		\item $\mathcal{P}_{s}$ increases with the increasing  $P_{\mathrm{total}}$, the decreasing $\sigma^2$, or the decreaing $\gamma_0$. First, the larger $P_{\mathrm{total}}$ or the smaller $\sigma^2$ leads to the larger $P_t$ (Referring to the SINR model in Eq.~\eqref{Eq-gamma-u0}), thus enhancing the received SINR at $u_i$ and resulting in a higher $\mathcal{P}_{s}$. Then, a larger $\gamma_0$ raises the decoding threshold, thereby lowering $\mathcal{P}_{s}$.
	\end{itemize}
	
	
	\textbf{Summary:} Overall, all simulation results match closely with the theoretical results, thus verifying the accuracy of our model. Meanwhile, the findings from our results can provide many valuable practical insights for network operators and system engineers to optimize parameter configurations in the PASS. For instance, the spatial trends observed in~\cref{Fig-Ps-u0} offer guidance for user deployment strategies aimed at maximizing the downlink successful transmission probability across different waveguide coverage areas. Additionally, the variations in $\mathcal{P}_{s}$ depicted in \cref{Fig-Ps-parameter} serve as a reference for configuring the PASS under diverse environmental conditions (e.g., room size, network status), enabling power-efficient operation.
	
	\section{Conclusion}\label{Sec-Con}
	We investigated downlink transmission performance in PA-assisted indoor immersive communications. Specifically, we first developed a 3D spatial model that accurately characterizes the distribution of users, waveguides, and PAs in a typical indoor scenario. We then evaluated the PA-assisted downlink link performance, i.e., STP, which captures the critical location-dependent relationships between system components while quantifying the impacts of key deployment parameters. Extensive Monte Carlo simulations were conducted to verify the accuracy of the proposed theoretical model. In our future work, we will further unlock the potential of PA-assisted indoor immersive communications in addressing practical challenges. These include the integration of multiple access mechanisms, the mitigation of random blockage effects, and the management of multipath interference. How to develop comprehensive models and solutions for these challenges also represents an important direction for our future research.
	
	{\appendices
		\section{}
		\label{proof_pro1}
		\vspace{-0.3em}
		\textit{Proof of Proposition 1:} Substituting \cref{Eq-gamma-u0,Eq-Power,Eq-Interference} into $\mathbb{P}\left(\gamma_u^i > \gamma_0 \right)$, we have
		\begin{equation}
			\begin{aligned}
				\mathcal{P}_{s}
		& \overset{(a)}{=} \resizebox{0.8\hsize}{!}{$ \mathbb{P} \left( \frac{ ((id-y_u^i)^2 + h^2)^{-1} }{ \eta  \sum_{k \in \mathcal{K}/\{i\}} ((x_u^k-x_u^i)^2+(kd-y_u^i)^2 + h^2)^{-1} + \frac{\sigma^2}{P_t}} \geq \gamma_0 \right) $} \\
		& \overset{(b)}{=} \mathbb{P}\left( R \leq \frac{((id-y_u^i)^2 + h^2)^{-1}}{\eta \gamma_0} - \frac{\sigma^2}{\eta P_t} \right) \\
		& = F_{R} \left(z\right)|_{z= \frac{((id-y_u^i)^2 + h^2)^{-1}}{\eta \gamma_0} - \frac{\sigma^2}{\eta P_t}}, 
		\label{Eq-STP}
	\end{aligned}
\end{equation}
where $(a)$ is resulted from $x_p^i=x_u^i,x_p^k=x_u^k$ (because the PA's location has the same $x-$axis value with its user) and $y_p^i=id,y_p^k=kd$ (due to the deployment of waveguides) and $(b)$ comes from the substitutions of $R = \sum_{k \in \mathcal{K}/\{i\}} r_k$ and $r_k = ((x_u^k - x_u^i)^2 + (kd - y_u^i)^2 + h^2 ) ^{-1}$. 
\hfill $\blacksquare$

\section{}
\label{proof_the1}
\vspace{-0.3em}
\textit{Proof of Theorem 1:}
Let $\phi_{R}(t)$ denote the characteristic functions of $R$, the CDF $F_{R} \left(z\right)$ 
can be obtained by~\cite{ghatak2021stochastic}
\begin{align}
	&F_{R}(z)
	= \frac{1}{2} - \frac{1}{\pi} \int_0^{\infty} \frac{\mathrm{Im}\left( e^{-jtz} \phi_{R}(t) \right)}{t} \mathbf{d}t,
	\label{Eq_FR}
\end{align}
where $\mathrm {\mathrm{Im}}(\cdot)$ calculates the imaginary part. 
Recall that each PA's location is aligned with its served user and all users are independently distributed. Then the interfering PAs $\{p_k, k \in \mathcal{K}/\{i\}\}$ 
are independently and uniformly distributed. 
Let $\phi_{r_k}(t)$ denote the characteristic functions of $r_k$, we have 

\vspace{-0.5cm}
{\small
	\begin{align}
		&\phi_{R}(t) 
		= \phi_{\sum_{k \in \mathcal{K}/\{i\}} r_k}(t) 
		\overset{(a)}{=} \prod_{k \in \mathcal{K}/\{i\}} \phi_{r_k}(t),\label{Eq_PhiR}\\
		& \phi_{r_k}(t) 
		= \mathbb{E}_{r_k}\left[e^{jt r_k}\right] 
		\overset{(a)}{=} \mathbb{E}_{x_u^k}\left[e^{jt r_k} \right]
		\overset{(b)}{=} 
		\frac{1}{L} \int_{-L/2}^{L/2} e^{jt r_k} \mathbf{d}x_u^k,
		\label{Eq_Phir}
\end{align}}%
where \cref{Eq_PhiR}(a) comes from the independent characteristic of $\{r_k, k \in \mathcal{K}/\{i\}\}$. In \cref{Eq_Phir}, (a) comes from the variable $x_u^k$ in $r_k$, (b) is resulted from the uniform distribution of $x_u^k\in[-L/2,L/2]$ in the service region $\mathcal{A}_k$ of the waveguide $w_k$, where probability density function (PDF) of $x_u^k$ is $f_{x_u^k}(x_u^k)=1/L$, and (c) is due to the identical distribution of $\{x_u^k, k \in \mathcal{K}/\{i\}\}$. 
Substituting \cref{Eq_Phir,Eq_PhiR,Eq_FR} into \cref{Eq-STP}, 
we have
\vspace{-0.3cm}

{\footnotesize
	\begin{align*}
		&\mathcal{P}_{s}
		= \frac{1}{2} - \frac{1}{\pi} \int_0^{\infty} \mathrm{Im}\left( e^{-jtz} \prod_{k \in \mathcal{K}/\{i\}} \phi_{r_k}(t) 
		\right)  \frac{\mathbf{d}t}{t} \notag\\
	&\overset{(a)}{=} \frac{1}{2} - \frac{1}{\pi} \int_0^{\infty} \\
	&\mathrm{Im}\left( e^{-jtz} 
	\left(\frac{1}{L}\right)^{2K} 
	\underbrace{\int_{-\frac{L}{2}}^{\frac{L}{2}} \dots \int_{-\frac{L}{2}}^{\frac{L}{2}}}_{2K} e^{jt \sum_{k \in \mathcal{K}/\{i\}}r_k} \underbrace{\mathbf{d} \dots \mathbf{d}x_u^k}_{k \in \mathcal{K}/\{i\}} 
	\right) \frac{\mathbf{d}t}{t}\notag\\
	&\overset{(b)}{=} \frac{1}{2} - \frac{1}{\pi} \left(\frac{1}{L}\right)^{2K} \int_0^{\infty} \mathrm{Im}\left(  
	\underbrace{\int_{-\frac{L}{2}}^{\frac{L}{2}} \dots \int_{-\frac{L}{2}}^{\frac{L}{2}}}_{2K} e^{jt A} \underbrace{\mathbf{d} \dots \mathbf{d}x_u^k}_{k \in \mathcal{K}/\{i\}} 
	\right) \frac{\mathbf{d}t}{t},
\end{align*}%
}%
where $(a)$ is the independent distribution of $x_u^k$ and $(b)$ comes from the substitution of $A=-z+\sum_{k \in \mathcal{K}/\{i\}}(r_k )$. 
\hfill $\blacksquare$
}

\bibliographystyle{IEEEtran}
\bibliography{IEEEabrv, references.bib}
\end{document}